\documentclass[a4paper,
               biblatex,     
               keeplastbox,   
               nospread,     
               ]{jacow}
\usepackage{pdfpages,multirow,ragged2e} %
\usepackage[unicode=true]{hyperref}
\newcommand{\todo}[1]{\textbf{todo}}
\newcommand{\etal}{et~al~}
\makeatletter%
	\ifboolexpr{bool{xetex}}
	 {\renewcommand{\Gin@extensions}{.pdf,%
	                    .png,.jpg,.bmp,.pict,.tif,.psd,.mac,.sga,.tga,.gif,%
	                    .eps,.ps,%
	                    }}{}
\makeatother

\ifboolexpr{bool{xetex} or bool{luatex}} 
 {}                                      
 {\usepackage[utf8]{inputenc}}           

\usepackage[USenglish]{babel}

\ifboolexpr{bool{jacowbiblatex}}%
 {%
  \addbibresource{THPS048.bib}
 }{}
\begin{document}

\title{eLog analysis for accelerators: \\status and future outlook}

\author{A. Sulc\thanks{asulc@lbl.gov}, T. Hellert, ALS, Berkeley Lab, Berkeley, USA\\
T. Britton, A. Carpenter, D. Lawrence, D. Lersch, D. McSpadden, C. Tennant, \\Thomas Jefferson National Accelerator Facility, Newport News, USA\\
D. Ratner, A. Reed, C. Bisegni, A. Bien, SLAC National Accelerator Laboratory, USA\\
H. Hoschouer, J. St. John, Fermi National Accelerator Laboratory, USA\\
}
\maketitle

\begin{abstract}
This work demonstrates electronic logbook (eLog) systems leveraging modern AI-driven information retrieval capabilities at the accelerator facilities of Fermilab, Jefferson Lab, Lawrence Berkeley National Laboratory (LBNL), SLAC National Accelerator Laboratory. We evaluate contemporary tools and methodologies for information retrieval with Retrieval Augmented Generation (RAGs), focusing on operational insights and integration with existing accelerator control systems.

The study addresses challenges and proposes solutions for state-of-the-art eLog analysis through practical implementations, demonstrating applications and limitations.
We present a framework for enhancing accelerator facility operations through improved information accessibility and knowledge management, which could potentially lead to more efficient operations.
\end{abstract}

\section{Introduction}
eLogs are essential for documenting accelerator operations, capturing everything from routine maintenance to complex beam studies. 
However, the volume of these records—spanning decades, multiple facilities, and varying formats makes retrieving relevant information difficult. This paper shows the current status of AI-driven solutions implemented at four accelerator facilities (Fermilab, Jefferson Lab, LBNL, and SLAC). 

\section{Related Work}
In the development of eLogs for accelerators and tokamaks, there are three notable works. First, Mayet~\cite{mayet2024gaia} implemented a multi-expert RAG system that combines LLMs with particle accelerator control systems and documentation tools using ReAct prompting~\cite{yao2023react}, enabling operators to retrieve expert knowledge, interact directly with the machine, and generate control scripts to simplify accelerator operations.
Second, Metha~\etal{}~\cite{mehta2023towards} a RAG system that leverages operator text logs from tokamak fusion experiments (DIII-D and Alcator C-Mod) to create an AI copilot that demonstrates superior performance over baseline methods in semantic search and device-specific operations assistance.
Lastly, Maldonado~\etal{}\cite{maldonado23enhancing} used OCR to use often present images in the BNL logbook entries and used them to augment the existing entries. Recently, Sulc~\etal{}~\cite{sulc2024towards}~surveyed the eLog activities.

\section{Current status}
\subsection{Advanced Light Source}

\begin{figure}[ht!]
\includegraphics[width=1\linewidth]{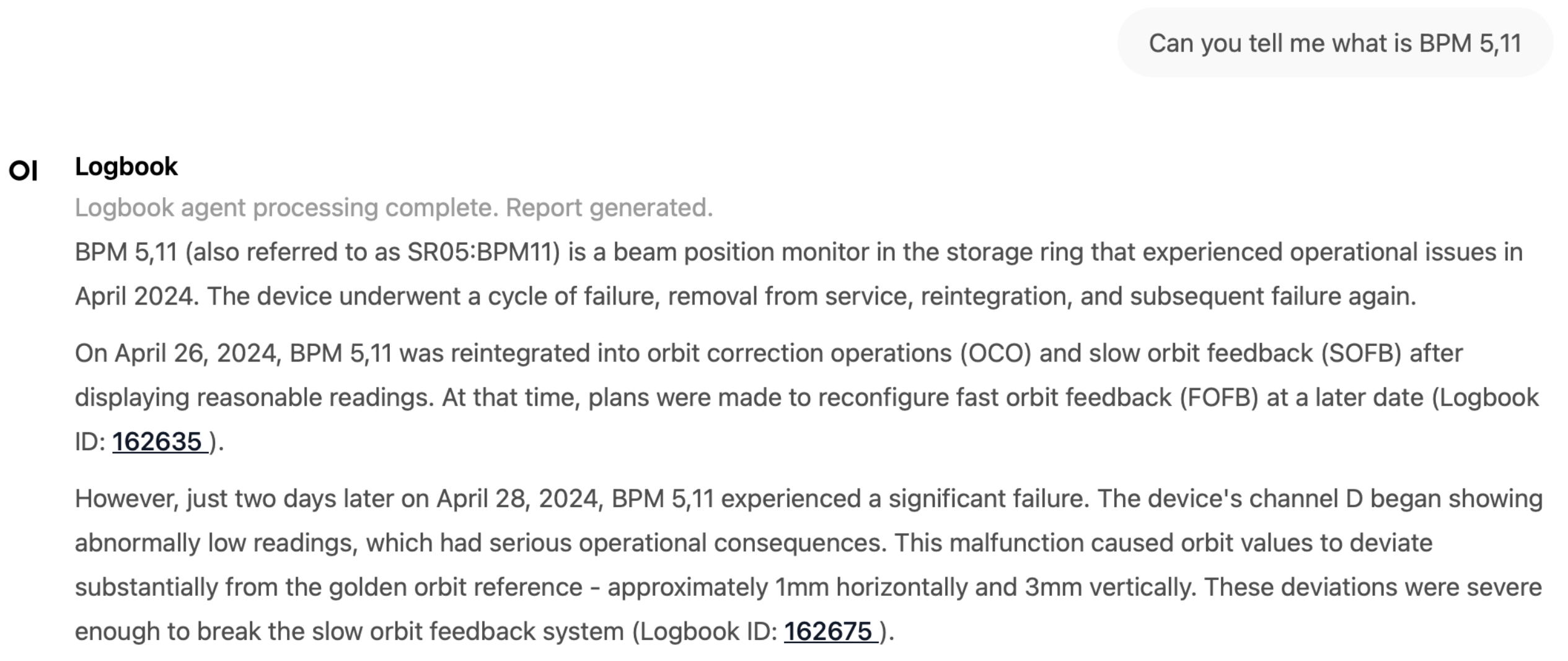}
\caption{Screenshot of the currently developed system built on OpenWebUI~\cite{openwebui2024} interface.}
\label{fig:ALS-LOGBOOK}
\end{figure}

The ALS at LBNL utilizes an logbook system integrated as an Model Context Protocol (MCP) client within its accelerator control infrastructure, see Fig. \ref{fig:ALS-LOGBOOK}. 

The ALS eLog system is integrated as an MPC client that embeds RAG as a functional component within the broader control framework. 
Information retrieval is performed using a fused approach combining \texttt{all-MiniLM-L6-v2} text embeddings and \texttt{BM25} keyword search scores, indexed and queried through a Qdrant vector database. 
Since embedding and keyword search use different scaling, we used reciprocal rank. 
While recognizing the potential of superior embedding models, the current system relies on \texttt{all-MiniLM-L6-v2}, with plans underway to integrate a custom-finetuned \texttt{AccPhysBERT} model.

After an initial high-recall search using fast vector embeddings (which only requires embedding the input query and performing matrix multiplication), we apply a re-ranking step to improve precision. While re-ranking is computationally more expensive by requiring embeddings of all filtered entries, it delivers entries with better accuracy on already prefiltered entries and refines the search. 
This stage employs the \texttt{cross-encoder/ms-marco-MiniLM-L-2-v2} model for efficient filtering, discarding entries below an experimentally determined score threshold 
(currently >0). 
The filtered, relevant entries are then processed by an LLM for summary generation. Operational experience and feedback from ALS personnel suggest that near-term improvements in retrieval accuracy are prioritized for enhanced utility, as the summarization capabilities, especially with smaller language models, currently exhibit limitations related to hallucinated content.

\subsection{Fermilab}
Since 2013, Fermilab has relied on ADEL (Accelerator Division Electronic Logbook) to capture all aspects of accelerator operations. ADEL is used to document machine failures, repairs,tuning, beam studies, as well as work planning and safety approvals~\cite{hazelwood_wao_2023}. It's comprehensive collection of operations activities makes ADEL a valuable resource for troubleshooting, allowing operators to reference past issues, resolutions, and patterns across the accelerator complex. 

\subsubsection{\textbf{ADEL Smart Search}}
To improve how operators retrieve past entries, a semantic search tool—Smart Search—has been deployed in the Fermilab AD Main Control Room. Hosted on the AD Kubernetes cluster, Smart Search enables two modes of retrieval: written queries and a recommendation system that finds similar notes based on an existing logbook entry.
Each ADEL entry is converted into a 768-dimensional vector using the SBERT model~\texttt{all-mpnet-base-v2}~\cite{reimers-2019-sentence-bert}, then stored in a Qdrant vector database~\cite{Qdrant}. 
Qdrant also stores metadata alongside each vector—such as creation date, machine type, and subsystem category (e.g. Linac, RF, Tuning)—enabling precise filtering.
Smart Search builds on Qdrant’s native recommendation support, which allows the use of both positive (similar) and negative (dissimilar) examples to refine search results. 
Whether using a query or a reference entry, the system retrieves semantically similar entries based on vector similarity. With filters applied, operators can quickly isolate related incidents or resolutions tied to specific systems without relying on exact phrasing or manual guesswork.

\subsubsection{Entry Augmentation}
Fermilab operates several accelerators that share similar subsystem labels. For example, both the Booster and Linac have RF and Water systems. While domain experts can easily distinguish between these based on context, embedding models often struggle to make such distinctions without additional guidance. To address this limitation, each logbook entry was augmented with contextual metadata prior to vectorization. This metadata—such as machine and subsystem labels—is manually assigned by the user at the time of entry creation. 

During preprocessing, this information was added to the beginning of each entry in an NLP-compatible format. For example, an entry tagged as originating from the Booster and involving RF would include statements such as “This is related to Booster” and “This involves RF.” This augmentation improved semantic clustering and led to more accurate retrieval of relevant entries, particularly in cases where the original text did not explicitly state the associated machine or subsystem.

\subsubsection{Adaptive Thresholding for Retrieval}
While semantic search is effective for static knowledge bases, applying it to a logbook presents a unique challenge: logbooks are collections of time-stamped operational events. An entry may be semantically similar to a user’s query—for example, a note on tuning Linac—but if it was recorded six years ago, its practical relevance may be limited. In this context, temporal proximity can be just as important as textual similarity.
To address this, the number of retrieved entries was set deliberately high during the initial search phase. An adaptive similarity threshold was then applied to exclude low-relevance results while preserving a broad set of meaningful matches. This threshold was determined by analyzing the distribution of similarity scores and applying heuristics to identify a natural cutoff point. This allowed the remaining entries to be sorted by time of entry creation without semantically weak or unrelated entries disrupting the ranking. As a result, operators could more effectively prioritize recent, contextually relevant notes without compromising retrieval quality.

\subsection{Jefferson Lab}
Jefferson Lab is pursuing several complementary research directions to integrate LLMs with the Continuous Electron Accelerator Facility (CEBAF) logbook, each addressing a distinct challenge in turning years of operational records into an AI-ready knowledge base:
\subsubsection{Unified Formatting}
To feed log entries into LLMs, we first normalize and convert their HTML or plain-text forms into clean Markdown~\cite{tennant2025logbook}. By standardizing formats - including tables and embedded images - we
ensure that downstream LLM pipelines can reliably parse and reason over each record.

\subsubsection{Domain Ontology}
LLMs excel at pattern recognition but struggle with highly specialized domains. We are constructing an explicit ontology of CEBAF-specific concepts, defining each term’s properties and interrelations in a machine-interpretable graph. This ontology augments semantic search by driving synonym expansion and disambiguation (e.g., linking "cavity" to its RF-accelerator meaning), and informs RAG prompt templates, ensuring answers remain factually anchored. 

\begin{figure}[ht!]
    \centering
    \includegraphics[width=1.0\linewidth]{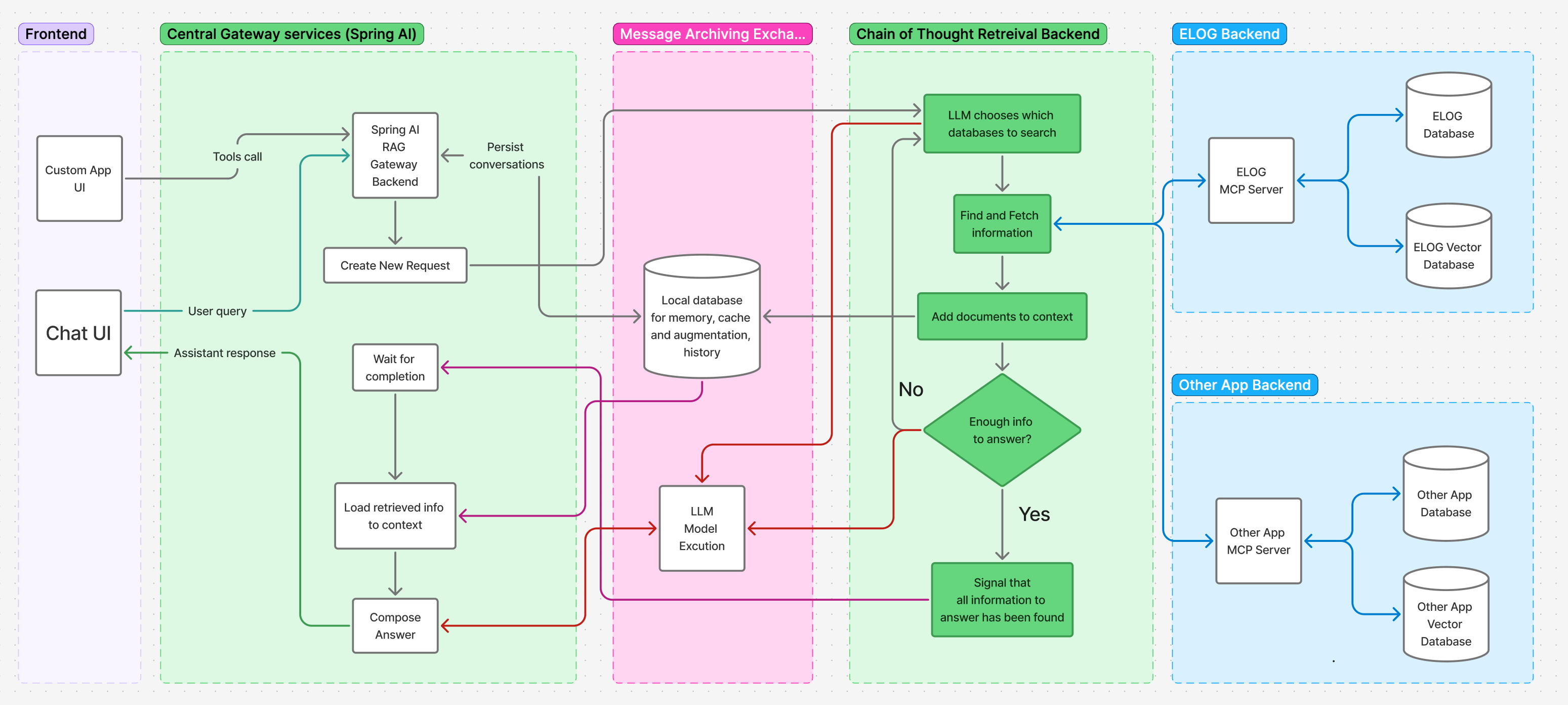}
    \caption{System architecture for SLAC AI tools.}
    \label{fig:slac}
\end{figure}

\subsubsection{Semantic Search}
Technical logbook entries are terse, abbreviation-heavy, and often blend narrative with tables or images, confounding off-the-shelf NLP pipelines. Using a specialized, multimodal semantic‐retrieval assistant, we convert each log entry into embedded “chunks” (including text, tables, and figures), index them in a vector database, and perform nearest-neighbor searches at query time. This framework preserves context across spelling variants, shorthand, and shifting nomenclature, vastly improving recall over legacy keyword search. Operators pose natural-language questions and can receive both improved search results and concise, cited answers
synthesized via RAG.  


\subsubsection{Enriched Log Entries} 
Human-authored log entries exhibit significant variance in detail and focus, reflecting differing expertise levels among contributors. When graduate students and specialized experts
document the same events, they naturally emphasize different aspects based on their knowledge domains. This inconsistency often results in omitted critical details. By leveraging
their understanding of CEBAF operational patterns and correlations across multimodal data sources, LLMs can enrich these human-created logs with supporting information, automatically
supplementing entries with relevant technical context and related data. With these building blocks – uniform formatting, ontology grounding, semantic search, and enriched log entries - we unlock a host of downstream capabilities.

\subsection{SLAC}

Scientists and operators at SLAC use electronic logbooks (ELOGs) to communicate with each other and record important information. Physicists at LCLS and FACET use the ELOG to record data, instrument settings, analysis results, scientific hypotheses, and shift summaries. Meanwhile, operators use logbooks to communicate about equipment status, repairs, diagnostic data, and logistics between sectors, in addition to shift summaries.

Physicists typically post unannotated screen grabs captured during experiments, making it difficult for other researchers to understand. For operators, creating shift summaries by hand takes significant time, and other posts are brief and lack tags. In both cases, it is often difficult to piece together what happened during a shift after the fact due to the fragmented, real-time nature of logbook entries.

To solve this, we are building an LLM-powered retrieval-augmented generation (RAG) system. (Fig. \ref{fig:slac}) User questions are embedded and matched via cosine similarity against the vectorized logbook; the retrieved entries are fed into the LLM, which composes an answer grounded in the original posts. The frontend uses Open WebUI, while a Spring AI orchestrator, Ollama on local GPUs, and a vector database form the containerized backend—ready for future Kubernetes deployment.

Upcoming enhancements include:
Hard filters (e.g. fixed date ranges), post interdependencies (e.g. amendments), cross-referencing external systems (ticketing, inventory), inline citations, and multimodal support (images and text). We also expose an API that auto-suggests titles and tags during drafting—promoting consistent metadata and improving later searchability.

Another approach leverages subject-matter expertise to continually update an internal wiki that acts as the RAG context base. This ensures that the wiki remains the single source of truth for all operational knowledge. By connecting SME-validated Q\&A directly to article sections geographically/hierarchically and linking them in a lightweight knowledge graph, the model retrieves context that is both current and accurate. As operators navigate from section to section, we log the paths taken (from question to final wiki article) to refine the orchestrator’s search strategies. Over time, this curated, graph‐structured wiki empowers our RAG system to convert individual operations ELOG entries into verbose micro-summaries that that roll up into fully automated, high-fidelity shift reports.

\section{Conclusion}
This work demonstrated practical RAG implementations across four accelerator facilities, each addressing facility-specific challenges through tailored approaches. 

Key technical contributions include: hybrid retrieval combining dense embeddings with BM25 scoring and MCP integration (LBNL), adaptive temporal thresholding with metadata augmentation for machine disambiguation (Fermilab), semantic search with ontology grounding and automated entry enrichment (Jefferson Lab), and API-driven auto-tagging with SME-validated wiki integration for shift report generation (SLAC). While current implementations show first use for semantic search and knowledge retrieval, challenges remain mostly in challenging input format with complex jargon or acronyms that lead to hallucinations. 

\section{Acknowledgments}
This document was prepared using the resources of the Fermi National Accelerator Laboratory, a U.S. Department of Energy, Office of Science, Office of High Energy Physics HEP User Facility. Fermilab is managed by FermiForward Discovery Group, LLC, acting under Contract No. 89243024CSC000002. Jefferson Lab's work was supported by the U.S. Department of Energy, Office of Science, Office of Nuclear Physics under Contract No. DE-AC05-06OR23177. ALS work was supported by the Director of the Office of Science of the U.S. Department of Energy under Contract No. DEAC02-05CH11231. Work at SLAC was supported by the U.S. Department of Energy, Office of Science, Basic Energy Sciences, under Contract No. DEAC02-05CH11231.

\ifboolexpr{bool{jacowbiblatex}}%
	{\printbibliography}%
	{%

}

\end{document}